# OS Scheduling Algorithms for Memory Intensive Workloads in Multi-socket Multi-core servers

Murthy Durbhakula
Indian Institute of Technology Hyderabad, India
cs15resch11013@iith.ac.in, murthy.durbhakula@gmail.com

*Abstract*—Major chip manufacturers have all introduced multicore microprocessors. Multi-socket systems built from these processors are routinely used for running various server applications. Depending on the application that is run on the system, remote memory accesses can impact overall performance. This paper presents a new operating system (OS) scheduling optimization to reduce the impact of such remote memory accesses. By observing the pattern of local and remote DRAM accesses for every thread in each scheduling quantum and applying different algorithms, we come up with a new schedule of threads for the next quantum. This new schedule potentially cuts down remote DRAM accesses for the next scheduling quantum and improves overall performance. We present three such new algorithms of varying complexity followed by an algorithm which is an adaptation of Hungarian algorithm. We used three different synthetic workloads to evaluate the algorithm. We also performed sensitivity analysis with respect to varying DRAM latency. We show that these algorithms can cut down DRAM access latency by up to 55% depending on the algorithm used. The benefit gained from the algorithms is dependent upon their complexity. In general higher the complexity higher is the benefit. Hungarian algorithm results in an optimal solution. We find that two out of four algorithms provide a good trade-off between performance and complexity for the workloads we studied.

## 1 INTRODUCTION

Many commercial server applications today run on cache coherent NUMA (ccNUMA) based multi-socket multi-core servers. Depending on the application, DRAM accesses can impact the overall performance; particularly remote DRAM accesses. These are inherent to the application. One way to ameliorate this problem is to rewrite the application. Another way is to observe remote DRAM access patterns at run-time and adapt the OS scheduler to minimize the impact of these accesses. In this paper, we present such an OS scheduling optimization which observes local and remote DRAM accesses for each scheduling quantum and then applies one of the four different algorithms to decide where to schedule each thread for the next scheduling quantum. The main idea is to schedule a thread with most remote accesses coming from, say Node N, to that node N. For example, in a four node system, if thread T0 is currently scheduled on Node N0 and we observe that there are only 10 DRAM accesses to node N0, 10 to Node N1, 10 to Node N2, and 1000 to Node N3, then we schedule thread T0 to node N3 in the next scheduling quantum. To the best of our knowledge, none of the current commercial operating systems optimize their scheduler for remote DRAM accesses. We show that some of our scheduling algorithms can cut down DRAM accesses by up to 55% depending on the workload and DRAM latencies. In this study we assume each socket/node has 4 cores and each core can run one thread of a parallel application at-a-time. The rest of the paper is organized as follows: Section 2 discusses the scheduling algorithms. Section 3 describes the methodology I used in evaluating the scheduling algorithms. Section 4 presents results. Section 5 describes related work and Section 6 presents conclusions.

## 2 SCHEDULING ALGORITHMS

Various scheduling algorithms can help reduce remote memory accesses. We consider four different algorithms in this paper. Three of them are greedy algorithms and fourth one is based on the Hungarian algorithm. For all the algorithms we assume dedicated hardware performance counter support to count local and remote memory accesses.

For every thread there are as many counters as the number of nodes in the system. In each scheduling quantum for every read request to DRAM we keep track of which thread made the request and whether the access is local DRAM access or remote access and increment corresponding counter. At the end of the scheduling quantum OS reads all the performance counters and applies one of the four scheduling algorithms and makes scheduling decision for the next quantum. We describe these algorithms in the next sub-section. While describing the algorithms we assume a four-node system with each node capable of running four different threads on four cores. Hence the system can run a total of 16 threads.

### 2.1 Algorithm 1

In this algorithm, we first sort all local and remote memory access counts for each thread in monotonically decreasing order. We start from top and assign the thread with highest access count to that node. Say thread T0 has highest access count of 10000 to node 2 then we assign T0 to node 2. Then we go to next element and assign the next thread to its corresponding node. Once we assign a thread to one node we cannot assign it to another node. So in the above example once T0 is assigned to node 2, it cannot be assigned to any other node. Further once we reach a maximum of four threads assigned to a particular node then we no longer assign any more threads to that node and simply skip assigning more threads to that node. This is for load balancing reasons. The complexity of the algorithm is of order $O(NLlog(NL))$ where N is total number of threads and L is total number of nodes. This is because the sorting algorithm dominates the complexity and we use merge-sort. There are specialized algorithms such as counting sort which could further reduce complexity, however their application is data dependent.

---

Input: Threads T0,…TN with DRAM accesses to nodes N0…NL. Each node has 4 cores and can run one thread per core. Current schedule S_current which has a mapping of N threads to L nodes.

Output: New schedule S_next with new mapping of threads to nodes.

begin

1. Each of N threads have L DRAM access counts; one per node. Sort N*L DRAM accesses in descending order. Complexity is $O(NLlog(NL))$.

2. Scan DRAM access counts. Start from highest DRAM access count and say it is coming from thread T1 to node N2. Assign T1 to N2 in schedule S_next. Now T1 cannot be assigned to any other node.

3. Repeat step 2 until all threads are assigned a node in S_next. Complexity is $O(NL)$.

end

Overall complexity of the algorithm is $O(NLlog(NL))$.

---

**Algorithm 1**

## 2.2 Algorithm 2

In this algorithm we first start off with node 0 and we sort all thread accesses to that node in monotonically decreasing order. Then we pick first four threads with highest accesses to node 0. Then we remove those four threads from the list. We then sort thread accesses to node 1 out of 12 remaining threads. We pick top four threads to node 1. We remove these four threads from the list. We sort accesses to node 2 from the remaining 8 threads. We pick top 4 to node 2. Remaining 4 will go to node 3. The complexity of this algorithm is order $O(Nlog(N))$ assuming sorting is done in parallel on all nodes, where N is total number of threads. Even here it is sorting that dominates the complexity.

---

Input: Threads T0,…TN with DRAM accesses to nodes N0…NL. Each node has 4 cores and can run one thread per core. Current schedule S_current which has a mapping of N threads to L nodes.

Output: New schedule S_next with new mapping of threads to nodes.

begin

1. for(i=0; i<L;i=(i+1))

    begin

       Sort all DRAM accesses from N threads

       in descending order

    end

    If all nodes do it in parallel. Complexity is $O(Nlog(N))$.

2. Start with node N0 and pick top 4 threads and assign them to N0 in schedule S_next. Now these threads cannot be assigned to any other node.

3. Repeat step 2 for all nodes until all threads are assigned a node in S_next. Complexity is $O(N)$.

end

Overall complexity of the algorithm is $O(Nlog(N))$.

---

**Algorithm 2**

## 2.3 Algorithm 3

Unlike previous two algorithms here we form combinations of all four threads and their summation of local accesses to each node. These algorithms are then sorted in monotonically decreasing order. We start by picking the topmost combination of four threads. We then remove all combinations where any of these threads occur from our list. Then the topmost combination from remaining elements is picked and scheduled on corresponding node. This process is repeated until all threads are scheduled. The complexity of this algorithm is clearly more than the previous two. It is O(L*N©4 log (L*N©4)) where N©4 is N combinatorial 4, N being total number of threads and L being total number of nodes. This is because of the combinations of four threads we form from N threads and for sorting of those combination

---

Input: Threads T0,…TN with DRAM accesses to nodes N0…NL. Each node has 4 cores and can run one thread per core. Current schedule S_current which has a mapping of N threads to L nodes.

Output: New schedule S_next with new mapping of threads to nodes.

begin

Compute all combinations of group of 4 threads from N threads. Complexity is O(N©4) where N©4 is N combinatorial 4.

For each combination of 4 threads compute total DRAM access count for every node. Complexity is O(L*N©4).

Sort the combinations obtained in step 2 in descending order. Complexity is O(L*N©4log(L*N©4)).

Scan the sorted list created in step 3 starting from top. Assign the group of 4 threads with highest count to corresponding node. Once these threads are assigned they can no longer be assigned to any other node. Keep scanning and assigning threads to nodes until all threads are assigned to nodes in S_next. Complexity if O(L*N©4).

end

Overall complexity is o(L*N©4log(L*N©4))

---

**Algorithm 3**

### 2.4 Algorithm 4

Instead of combinatorial enumerations, we can model the problem as an assignment problem for which a more optimal polynomial time Hungarian algorithm can be applied. The complexity of optimized Hungarian algorithm is known to be $O(N^3)$ which is better than that of algorithm 3. For each of the N threads there are N places across L nodes where they can be assigned. The Hungarian algorithm finds optimal placement of threads for which overall DRAM access latency (both for local and remote accesses combined) is minimized. I am not showing pseudo-code here as it can be found in many places. Original Hungarian algorithm can be found here [10].

## 3 METHODOLOGY

We implemented each of these algorithms as a stand-alone C++ program and evaluated them by running synthesized data access patterns. These synthesized data access patterns vary the local/remote dram access counts in a known pattern for each scheduling quantum while the counts themselves are generated as a random number in the range 0 to 10000 accesses. The data is then fed to each of the algorithms to see if they can track the pattern. The overall benefit for each algorithm depends on how well they can track the pattern. We chose synthetic workloads in order to clearly bring out benefits of each algorithm and its sensitivity to remote DRAM latency. As part of future work we plan to study the impact of these algorithms on real workloads. Further in this study we do not take into consideration impact of other parameters such as cache-to-cache transfers, cache-affinity, etc. while coming with a new schedule. We purely focus on optimization for remote DRAM latency.

Table 1 lists base configuration used for evaluation of different scheduling algorithms. On top of the base configuration we also perform sensitivity analysis of different algorithms with respect to varying remote DRAM latency from 150 to 200 to 300 cycles. Table 2 lists three synthetic DRAM access patterns used in this paper. A high level system diagram can be seen in Figure 1. The circled portion represents one socket with 4 cores per socket.

| Parameter | Value |
| --- | --- |
| CPU Frequency | 1 Ghz |
| Number of cores per socket | 4 |
| Number of sockets/nodes | 4 |
| Local DRAM Latency | 100 cycles |
| Remote DRAM Latency | 150 cycles |
| Number of scheduling quanta | 16 |

**Table 1: Configuration Parameters**

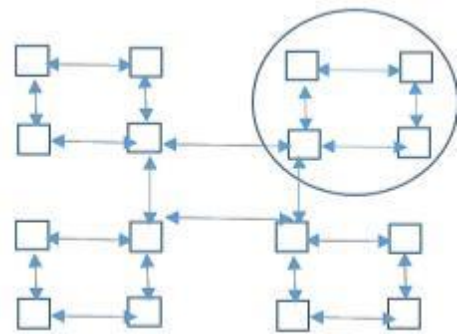

**Figure 1: High-level ccNUMA Multi-socket Multi-core System diagram with 4 sockets and 4 cores-per-socket**

| Access Pattern | Description |
|---|---|
| Constant access pattern – Synth1 | Same access pattern except for the first quanta. Single optimum grouping for all 16 quanta except for the first one. |
| 2-phase access pattern – Synth 2 | Two access patterns equally distributed over all quanta except for the first one. Optimal grouping 1 till $9^{th}$ quanta following by optimal grouping 2 till $16^{th}$ quanta |
| 4-phase access pattern – Synth 3 | Four access patterns equally distributed over all quanta except for the first one. Optimal grouping 1 till $5^{th}$ quanta followed by optimal grouping 2 till $9^{th}$ quanta followed by optimal grouping 3 till $13^{th}$ quanta followed by optimal grouping 4 till $16^{th}$ quanta. |

**Table 2: Synthetic Workloads**

## 4 RESULTS

All three workloads were run on four algorithms described in Section 2. Table 3 shows percentage benefit from each algorithm in terms of number of DRAM cycles saved for all sixteen threads running on 4 sockets for 16 scheduling quanta.

| Workload | Algo 1 | Algo 2 | Algo 3 | Algo4 |
|---|---|---|---|---|
| Synth1 | 13 | 16 | 25 | 25 |
| Synth2 | 13 | 15 | 23 | 23 |
| Synth3 | 9 | 11 | 19 | 19 |

**Table 3: Performance comparison of different scheduling algorithms (in % DRAM cycles saved). Remote DRAM latency 150 cycles.**

As we can see from Table 3, for each algorithm there is a consistent decrease in benefit as we move from workload 1 to 2 to 3. This is expected because as we move from workload 1 to 3 there is more variance in DRAM access patterns. And since we use past behavior as an indication of future all the algorithms would have, in general, greatest benefit for workloads having less variance.

Further we also see that Algorithm 3 and Algorithm 4 has highest benefit compared to other algorithms. This is because they are more sophisticated compared to other algorithms. Algorithm 4 gives as much benefit as algorithm 3 with added benefit that it is less complex and hence more scalable. Depending on the usage of the system one needs to decide which Algorithm to use. Algorithm 2 and Algorithm 4 seems to provide a good trade-off between performance and complexity.

We also performed sensitivity analysis of benefit of these algorithms under varying remote DRAM latency. We varied the latency from 150 to 200 to 300 cycles. Tables 4 and 5 shows results for remote DRAM latency of 200 and 300 cycles respectively.

| Workload | Algo 1 | Algo 2 | Algo 3 | Algo4 |
|---|---|---|---|---|
| Synth1 | 21 | 25 | 39 | 39.3 |
| Synth2 | 21 | 24 | 36 | 36 |
| Synth3 | 14 | 18 | 31 | 31 |

**Table 4: Performance comparison of different scheduling algorithms (in % DRAM cycles saved). Remote DRAM latency 200 cycles.**

| Workload | Algo 1 | Algo 2 | Algo 3 | Algo4 |
|---|---|---|---|---|
| Synth1 | 30 | 35 | 55 | 55.2 |
| Synth2 | 30 | 34 | 50 | 50 |
| Synth3 | 20 | 25 | 43 | 43 |

**Table 5: Performance comparison of different scheduling algorithms (in % DRAM cycles saved). Remote DRAM latency 300 cycles.**

As we can see from Table 4, Table 5, performance increases consistently in almost all cases as we increase remote DRAM latencies from 150 to 200 to 300 cycles. Similar to the base case we see that Algorithm 3 and Algorithm 4 has highest benefit compared to other algorithms. There is also a consistent decrease in benefit as we move from workload 1 to 2 to 3 due to reasons explained above. Algorithm 2 and Algorithm 4 again seems to provide a good trade-off between performance and complexity.

## 5 RELATED WORK

Chandra el al.[1] have studied impact of various OS scheduling policies on performance of both uniprocessor and multiprocessor workloads. Our work is similar to theirs in that we also study OS scheduling algorithms for improving performance of parallel workloads. However our focus more on memory intensive workloads for ccNUMA multi-socket multi-core servers. We assume that at-a-time only one multi-threaded application is running with each core assigned one thread for the sake of load-balancing. The algorithms we propose are completely different. Their work also evaluated benefits of page migration. However page migration & replication are not always beneficial. For instance say thread T0 is scheduled on Node 1 with 1000 accesses to Node 1's DRAM and also accesses a page P2 on Node 2 DRAM, with intention-to-write, 500 times. At the same time thread T1 is scheduled on Node 3 with 2000 accesses to Node 3's DRAM and accesses same page P2 on Node 2 DRAM, with intention-to-write, 500 times. In such a situation it is not possible to decide to which node to migrate P2 to. In fact if there are such "hot" pages accessed by multiple threads it is actually better to schedule them on the same node where hot page is, rather than migrating the hot page.

Kaseridis et al. [2] proposed a dynamic memory subsystem resource management scheme that considers both cache capacity and memory bandwidth contention in large multi-chip CMP systems. Their memory bandwidth contention algorithms monitor when a particular schedule of threads exceed the maximum bandwidth supported by a node and then try to schedule bandwidth demanding threads with those threads that need little memory bandwidth. Whereas our approach proactively tries to find the best pairing of threads for any scheduling quanta while keeping the overall bandwidth utilization within the maximum bandwidth limits. And the algorithms presented in this paper are different from their algorithms.

Ipek et al [3] proposed using reinforcement learning based approach to tune DRAM scheduling policies to effectively utilize off-chip DRAM bandwidth. Their work differs from ours in two different ways. First, we are using OS scheduling algorithms to optimize DRAM bandwidth utilization. Second, we use parallel workloads to evaluate the scheduling policy where as they use multiprogramming workloads with no sharing.

Ahn et al [4] studied the impact of DRAM organization on the performance of data parallel memory systems. In contrast to their work we focus on novel OS scheduling algorithms that will improve DRAM performance of parallel applications on general purpose multiprocessors. Zhu et al [5] proposed using novel DRAM scheduling algorithms for SMT processors. In contrast to their work this paper proposes using new OS scheduling algorithms with minimal hardware support.

Tang et al.[7] studied the impact of co-locating threads of different multi-threaded applications in a data-center environment on overall performance. They proposed heuristics for co-locating different workloads. Their focus is mostly on data-center related workloads and the algorithms presented in this work are different from their heuristics.

There are many other studies [6, 8, and 9] which focused on tuning DRAM scheduling policies or memory access ordering for better overall performance. Our work is different from these in two ways. First we focus on OS scheduling algorithms to reduce the impact of remote DRAM accesses. Second, these studies [6] focus on optimal DRAM utilization for co-located single threaded workloads where as our work focuses on improving performance of a multi-threaded parallel workload.

# 6 CONCLUSION

Many commercial server applications today run on ccNUMA multi-socket multi-core based servers. These applications typically suffer from remote DRAM accesses that diminish their overall performance. This paper presented an operating system scheduling optimization to ameliorate the performance impact of remote DRAM accesses. By observing local and remote DRAM accesses for various threads and incorporating that into the OS scheduling decision, we come up with a new schedule for the next scheduling quantum. We presented three new scheduling algorithms followed by an adaption of an existing Hungarian algorithm. Depending on the scheduling algorithm used performance benefit varied across different synthetic workloads. We also performed sensitivity analysis of these algorithms under varying remote DRAM latency. We showed that some of the algorithms can cut down DRAM access cycles by up to 55% depending on the workload used. The benefit gained from the algorithms is dependent upon their complexity. Higher the complexity higher is the benefit. Hungarian algorithm and algorithm 2 provide a good trade-off between performance and complexity for the workloads we studied.

This work could be extended in several ways. One way is to monitor the benefit resulting from the scheduling algorithms and if performance is reduced by applying these algorithms then the optimization can be turned off. The application behavior can still be monitored in the background to see if it enters a phase where it is beneficial to turn the scheduling optimization on again. Another way is to use machine learning algorithms to learn any kind of phase behavior among prior scheduling quanta and incorporate that into scheduling decision for the next quanta. In general for any long running application with stable patterns, hardware could provide feedback to the OS, which could in turn use that information to adapt its policies to benefit application performance.

**ACKNOWLEDGEMENTS**

I would like to thank Prof. Alan Cox of Rice University for initially discussing with me the concept of optimizing OS scheduling algorithms for improving the performance of various workloads. I would also like to thank various reviewers of this work for their comments and feedback. Finally I would like to thank my wife and kids for supporting me morally during the course of this work.